\documentclass[prl,twocolumn,preprintnumbers,amsmath,amssymb]{revtex4}

\usepackage{graphicx}
\usepackage{color}
\usepackage{bm}

\usepackage{amssymb, dsfont}

\newcommand{\R}{{\mathbf r}}

\newcommand{\f}{{\mathbf f}}
\newcommand{\F}{{\mathbf F}}
\newcommand{\T}{{\mathbf T}}

\newcommand{\M}{{\mathbf M}}
\newcommand{\K}{{\mathbf K}}
\newcommand{\I}{{\mathds 1}}

\newcommand{\E}{{\hat{\mathbf e}}}
\newcommand{\V}{{\mathbf V}}

\newcommand{\W}{\mbox{\boldmath$\Omega$}}

\newcommand{\beq}{\begin{equation}}
\newcommand{\eeq}{\end{equation}}
\newcommand{\bea}{\begin{eqnarray}}
\newcommand{\eea}{\end{eqnarray}}

\begin{document}

\title
{Self-Starting Micromotors in a Bacterial Bath}

\author{Luca Angelani$^1$, Roberto Di Leonardo$^2$, and Giancarlo Ruocco$^{2,3}$}

\affiliation{
$^1$Research center SMC INFM-CNR, c/o Universit\`a
di Roma ``Sapienza'', I-00185, Roma, Italy \\
$^2$Research center Soft INFM-CNR, c/o Universit\`a di
Roma ``Sapienza'', I-00185, Roma, Italy \\
$^3$Dipartimento di Fisica, Universit\`a di Roma ``Sapienza'',
I-00185, Roma, Italy
}

\begin{abstract}

Micromotors pushed by biological entities, like motile bacteria, constitute a
fascinating way to convert chemical energy into mechanical work at the
micrometer scale.  Here we show, by using numerical simulations, that a
properly designed asymmetric object can be spontaneously set into the desired
motion when immersed in a chaotic bacterial bath.  Our findings open the way to
conceive new hybrid microdevices exploiting the mechanical power production of
bacterial organisms.  Moreover, the system provides an example of how, in
contrast with equilibrium thermal baths, the irreversible chaotic motion of
active particles can be rectified by asymmetric environments.

\end{abstract}
\maketitle


Ensembles of animate organisms behave in a very rich and surprising way if
compared to inanimate objects, as atoms or molecules in a gas or a liquid.
Everyone has been amazed by the cooperative motion of birds in a flock, fishes
in a school or wildebeest in a herd \cite{couzin03,flock}.
Also at the micrometer scale elementary living organisms, like
bacterial cells, show an extraordinary variety of behaviors, such as collective
motions \cite{sakg07,lsdkw07,wl00,hsg05}, complex chemical-mediated motility or
chemotaxis \cite{chemo}, spatiotemporal patterns \cite{dom04}, self-organized
structures \cite{sperm}, biofilms formation \cite{biofilm}.  An important
peculiarity of animate organisms is the fact that they can be self-propelled,
using a variety of different mechanisms for this purpose \cite{swimm}.  Motile
cilia and turned flagella are two example of evolutionary tricks adopted by
living organisms to accomplish the hard task of swimming at low Reynolds number
\cite{pur76}.  One can think about such ensembles of organisms as  open
systems, with a net incoming flux of energy (provided by nutrients) stored and
converted into mechanical motion by irreversible processes happening inside the
cell body.  The resulting dynamics breaks time inversion symmetry so that
asymmetric environments can result in directed motions which, in equilibrated
Hamiltonian systems, would be forbidden by detailed balance \cite{gkca07,wan}.
A natural question then arises: is it possible to rectify such a non
equilibrium dynamics to propel microdevices?

Biological molecular motors constitute a fascinating mechanism to generate
motion at the nanoscale \cite{rondelez,liu}.  When larger, micron sized,
structures need propulsion the preassembled motor units found in unicellular
motile organism may offer several advantages over isolated proteins.  In a
recent experiment \cite{hmtu06,hmu05} bacterial driven micromotors have been
assembled by biochemically attaching motile bacteria to a microrotary motor.
Such procedures require the construction of narrow tracks to induce a
unidirectional binding of bacterial cells on to the moving rotor with a
consequent increased complexity in designs and limited number of working
bacteria. 

Here we numerically show that a properly designed asymmetric motor immersed in
a chaotic bacterial bath can be spontaneously set into the desired motion.  Our
numerical findings suggest the possibility to construct new opportunely shaped
microdevices able to exploit the propelling power of motile bacteria.

Spinning a bundle of helical flagella, bacteria like {\it E. coli}, may swim
along their body axis with speeds of order 10 body lengths per second
\cite{E_coli}.  Decorrelation of velocity may occur via four different
mechanisms: tumbling, Brownian motion, mechanical interactions and hydrodynamic
interactions.  The first mechanism is a spontaneous tumble produced by a
temporary reversal in the spinning direction of the flagellar motor
\cite{turner}. Brownian motion can also be effective in producing diffusion of
orientation and hence of propelling direction.  Interactions with other
bacteria can be mechanical, by direct contact, or hydrodynamic, via flow
currents produced by the swimming motions.  Trying to mimic the behavior of an
elongated {\it E. coli} cell with a minimal model we only retain the two most
effective mechanisms that are tumbling and mechanical interactions. Hydrodynamic
interactions, occurring only through dipole or higher order multipoles, turn
out to be effective only over short distances where mechanical interactions
between elongated bodies are much more effective in reorienting the bacteria.
We directly checked that including hydrodynamic interactions has a negligible
effect on the mean squared displacements and on its crossover from  ballistic
to diffusive regimes.

Each cell is represented by an instantaneous position $\R_i$ and an orientation
$\E_i$ pointing in the free swimming direction. The elongated hard body of the
cells (length $l$ and thickness $a$) is modeled by the sum of $p$ short range
repulsive potentials centered at equally spaced locations along the cell axis
$\R_i^\beta=\R_i+d^\beta\E_i$ with $\beta=1,p$
and $d^\beta=(l-a)(2\beta-p-1)/(2p-2)$.
The neighboring cells will then act on the ith cell with a
system of forces $\F_i^{\beta}$ applied at $\R_i^\beta$:
\begin{eqnarray}
&&\F_i^\beta=\sum_{j\neq i, \gamma} \f(\R_i^\beta-\R_j^\gamma)\\
&&\f(\R)=\frac{A\R}{r^{n+2}} \label{softeq}
\end{eqnarray}
\begin{widetext}
\begin{center}
\begin{figure}[htbp]
\includegraphics[width=0.96\textwidth]{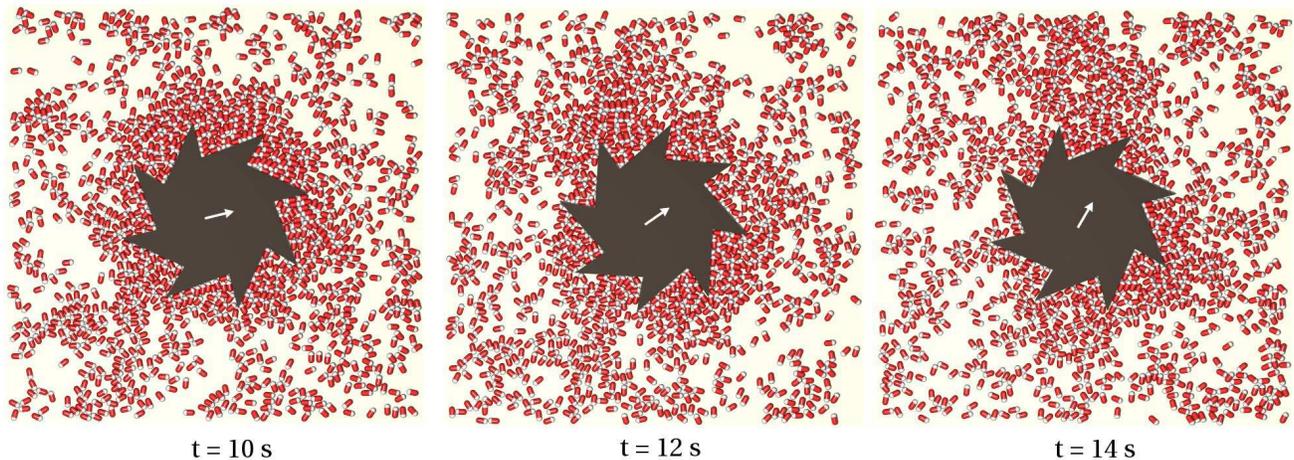}
\caption{(Color online) 
Rotary micromotor in a bacterial bath. 
Snapshots 
at three different simulation-times,
$t=10, 12$ and $14$ s. Each bacterium is represented by a spherocylinder 
(with aspect ratio $1/2$)
whit a white head pointing in the direction of the self-propelling force.
The arrow at the center of the gear evidences
the counterclockwise rotation at an average angular velocities 
$\omega_0 \simeq 0.21$ rad$/$s.
} 
\label{fig1}
\end{figure}
\end{center}
\end{widetext}
To such intercellular forces we added intracellular forces consisting of a
constant linear propelling force $f_0$ (directing along $\E_i$) which is only
active in the {\it running} state and a random torque $\T_r$ which switches on
during the {\it tumbling} state.  The probability per unit time to switch in a
tumbling state is constant and such as to give an average free run length of
$10$  cell lengths \cite{E_coli}.  
Introducing the state variable $\theta_i$ which is 0 in the
running state and 1 during a tumbling event, the net forces and torques acting
on the ith cell read: 
\begin{eqnarray} \F_i&=&f_0\E_i(1-\theta_i)+\sum_\beta\F_i^\beta\\
\T_i&=&\T_r\theta_i+\E_i\times\sum_\beta d^\beta\F_i^\beta \end{eqnarray}
For the subsequent motion the rigid cell body is modeled as a prolate spheroid
of aspect ratio $\alpha=a/l$.  Therefore the center of mass 
and the angular velocity are
\cite{kim}:
\begin{eqnarray}
\label{eom_bac1}
\V_i&=&\M_i\cdot \F_i\\
\W_i&=&\K_i\cdot\T_i
\label{eom_bac2}
\end{eqnarray}
where
\begin{eqnarray}
\M_i&=&m_{||}\E_i\E_i+m_{\perp}\left(\I-\E_i\E_i\right)\\
\K_i&=&k_{||}\E_i\E_i+k_{\perp}\left(\I-\E_i\E_i\right)
\label{Kmob}
\end{eqnarray}
We choose the force coefficient $A$ in such a way that two bacteria facing head to head
on the same line would be in equilibrium at a distance $a=\alpha l$:
\begin{equation}
A / a^{n+1}=f_0\Rightarrow A\simeq f_0 a^{n+1}
\end{equation}
We choose $l$ as the unit length, $\tau=l/v_0$ as the unit of time
(where $v_0=m_{||}f_0$ is the free swimming velocity) and $m_{||}$ 
as unit of mobility. 
When not specified  physical quantities will be 
expressed in reduced units.
A planar geometry will be investigated in a box $L \times L$ with periodic boundary conditions.
We will specialize to the case of $N=1092$ bacteria with
number density $\rho=N/L^2=0.945$, aspect ratio $\alpha=1/2$ and potential parameters
$p=2$, $n=12$. Mobility values are $m_{||}=1$, 
$m_{\perp}=0.87$, 
$k_{\perp}=4.8$ 
($k_{||}$ does not enter in the equation of motion, as $\T_i$ is perpendicular to
$\E_i$ in the planar geometry).
We consider a micromotor immersed in the bacterial bath. 
The asymmetric micromotor is a gear with a sawtooth profile whose 
center of mass is kept fixed at the center of the box.
The motor is free to rotate around its axis.
Each of the $p$ force centers, describing a single bacterium body, interacts with
boundary walls through a force of the form in Eq.~\ref{softeq}, where $\R$ is a vector
perpendicular to the wall connecting the $p$-centers to a point
located at a distance $a/2$ behind the wall.

\begin{figure}[t]
\begin{center}
\includegraphics[width=0.45\textwidth]{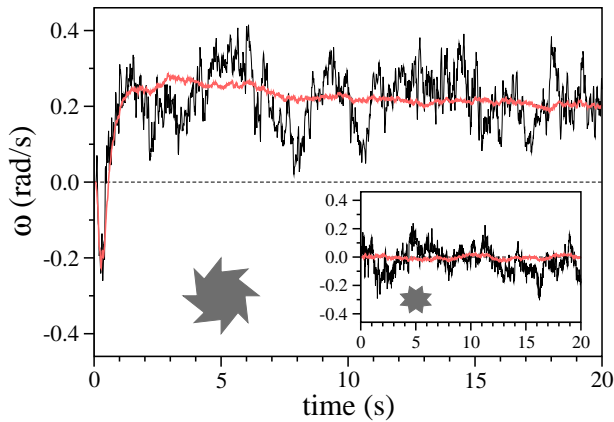}
\end{center}
\caption{(Color online)
Angular velocity $\omega$ (in rad/s) of the micromotor as a function of time (in seconds). 
Black-line refers to a single run, 
red-line is the average over 100 independent runs.
After a short transient regime (due to initial configuration of bacteria),
a fluctuating velocity around a mean value $\omega_0\simeq 0.21$ rad/s is observed.
Inset: same as main plot for a symmetrically shaped micromotor.
} 
\label{fig2}
\end{figure}
The resulting cell-boundary forces produce further force and torque terms 
in Eq.s~\ref{eom_bac1} and \ref{eom_bac2}, and a net 
fluctuating torque on the gear motor, whose angular velocity is then
\begin{equation}
\Omega_g = K_g\ T_g
\label{eom_mot}
\end{equation}
where $T_g$ is the torque exerted by bacteria on the gear whose
rotational mobility is $K_g$.
We consider a gear with $8$ teeth and internal (external) 
radius $R_{int}=5$ ($R_{ext}=8$).
The gear mobility is estimated as that of a disk \cite{kim} of radius $6.5$:
$K_g=1.9\cdot 10^{-3}$.
Equations of motion
(\ref{eom_bac1}), (\ref{eom_bac2}) and (\ref{eom_mot}) where numerically
integrated by Runge-Kutta method \cite{NR} for $2 \cdot 10^5$ steps (with time
step $\delta t=10^{-3}$).  At $t=0$ the bacteria are uniformly distributed in
the space outside the external disc or radius $r_{ext}$.

We found that the micromotor starts to move spontaneously under the effects of
pushing bacteria.  A net unidirectional motion is observed, with a fluctuating
angular velocity around a non-zero mean value.  In Fig.~1 we show snapshots of
the bacterial bath with a rotary micromotor at three different times, $t=10,
12$ and $14$ s  (physical units are obtained considering realistic values $l=3\
\mu$m, $v_0=30 \ \mu$m/s).  It is evident a densification process close to the
device's boundary, in agreement with recent studies on self-propelled cells in
confined environments \cite{conf1,conf2}.  As a result a net rotary
counterclockwise motion of the gear during time takes place.  The instantaneous
angular velocity $\omega$ of the motor as a function of time is shown in
Fig.~2, where  black-line refers to single run, while red-line is the
average over 100 independent runs.  After a short transient clockwise rotation
the system reaches a stationary regime with a fluctuating positive
(counterclockwise) angular velocity around a non-zero average value  $\omega_0
\simeq 0.21$ rad$/$s, corresponding to $2.0$ rpm .  The total torque on the
device can be estimated around $17$ pN $\mu$m 
(assuming $m_{||}=59$ $\mu$m/pN s).  
\begin{figure}[t] \begin{center} \hspace{1cm}
\includegraphics[width=0.55\textwidth]{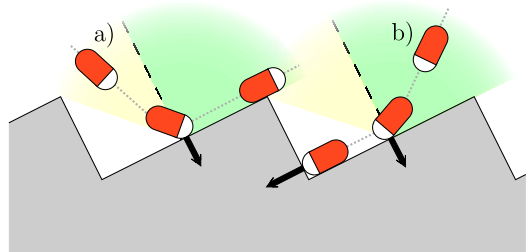} \end{center} \caption{(Color
online) Sketch of the collision of a single bacterium with the rotor boundary.
Arrows are the forces exerting by the bacterium onto the rotor.  a) Bacteria
coming from the left area with respect to the normal leave the gear.  b)
Bacteria from the right get stuck at the corner exerting a torque on the rotor.
} \label{fig3} \end{figure}
The onset of a directed rotation can be understood by analyzing the collision
of a single bacterium with the rotor boundary.  When a bacterium touches a
rotor edge it will exert a force given by the projection of the propelling
force onto the surface normal (black arrows in Fig.~3). The same force will act
on the bacterium producing a net torque that will align the cell body along the
edge.  Depending on the sign of the incident angle measured from the wall
normal, the bacterium will then quickly leave the gear back into the solution
(Fig.~3a) or get stuck at the corner exerting a torque on the rotor (Fig.~3b).
The same reasoning applies for both the long and short edges. Most of the
collisions, however, will occur on the long edge contributing a transient
negative torque which explains the short times negative dip in rotor angular
velocities.  It is worth noting that the elongated form of the bacteria is not
essential for the observed effect, as the same directed motion of the micromotor
also occurs in the presence of ``spherical'' bacteria, i.e. with aspect ratio
$\alpha=1$. The shape of the motor, instead,  plays a crucial role. Indeed,
simulations performed with a symmetric gear (with symmetrically shaped teeth)
produce on average an immobile motor, whose angular velocity fluctuates around
zero (inset in Fig.~2).  The asymmetry is then a basic ingredient, as observed
in many other thermal ratchet mechanisms discussed so far in the literature
\cite{flop,reim,note}.

Given the above mechanism, one expects that the torque $T_g$ exerted by
bacteria would increase as the square of the size $R$ of the rotor as both
perimeter (and hence applied forces) and moment arm increase linearly.  On the
other hand, the rotational mobility of the gear $K_g$ decreases as $1/R^3$
resulting in an average angular velocity decreasing as $1/R$. The maximal work
that can be extracted from the bath is obtained when an external reversible
system applies an opposing torque equal to $T_g/2$. The extracted mechanical
power is then given by $T_g^2 K_g/4$ and increases with $R$.  Therefore in a
planar geometry,  a 2D array of small gears would perform better, in terms of
usable power, than a single big one.  The dependence on bacterial concentration
is also non trivial due to the interbacterial interactions that could result in
reduced motilities at high packing fractions.  We note that, however, the
observed directed motion of the rotor is a quite robust effect with respect to
the variation of different physical parameters, as the density of bacteria,
their aspect ratio, the shape of the asymmetric rotor, its size, the boundary
conditions: a quantitative discussion on the role of different parameters will
appear in a forthcoming paper.

Our main point here is to demonstrate that, in contrast to thermal baths of
passive particles, useful work can be extracted from the chaotic motion of a
non-equilibrium suspension of active objects.  This behavior reminds of the
ratchet effect or Brownian motors \cite{reim}, in which out-of-equilibrium
systems undergo a rectification process in the presence of some asymmetric
potential or device.  More specifically, in an equilibrated Hamiltonian system,
there's no entropy production and time reversal symmetry guarantees that any
trajectory has the same probability than its time reversed, so that no
systematic directed motion can be observed on average.  On the other hand when
a self propelled particle collides to another (or to a boundary), the forces
they exchange is not just the repulsion of their rigid bodies, but there are
also the forces generated by the propelling units. Such forces are directed
along the incoming directions of the two particles and therefore would change
sign upon time reversal, while particles repulsion wouldn't. Time reversed
trajectories are then incompatible with the assumed dynamical laws.
From a thermodynamic viewpoint such irreversible dynamics
reflects the constant entropy production involved in the chemico-physical
processes driving the propelling unit, such as the flagellar rotary motor of
E-coli.
Once time inversion symmetry does not hold 
a spontaneous directed motion is allowed whenever a spatial 
inversion symmetry is broken \cite{note}.

In conclusion, we have shown that it is possible to conceive opportunely shaped
microdevices that can move in a directional way when immersed in a bath of
motile microorganisms. In particular,
we numerically show that a rotary micromotor, 
consisting of an asymmetric gear in a bath of {\it E. Coli} bacteria,
spontaneously sets into a unidirectional rotational
motion at an average speed of a few rpm.
Using asymmetrically shaped boundaries
also linear translatory motions could be obtained and bacterial driven
transport could be achieved by self assembly of bacteria along the particle's
boundary.  Remarkably, when coupled to an external reversible device, a net
amount of useful energy could be extracted from the chaotic motion of a
bacterial bath.  Our findings can open the way to new and fascinating
applications in the field of hybrid bio-microdevices engineering, and also
provide new insight in the more fundamental aspects of non-equilibrium dynamics
of active matter.

We acknowledge support from the INFM-CNR CINECA initiative for parallel computing.




\begin{thebibliography}{99}
\bibitem{note}
We note that a ratchet effect can also be 
observed in symmetric systems, due to a spontaneous symmetric breaking 
phenomenon \cite{dermeer}.
\end{thebibliography}

\end{document}